\documentclass[reprint, aps,prx,showpacs,bibfootnote,preprintnumbers, mathtools, superscriptaddress, amsmath,amssymb]{revtex4-2}

\usepackage{amsmath,amsfonts,amsthm,bm}
\usepackage{braket}
\usepackage[caption=false]{subfig}
\usepackage{graphicx}
\usepackage{xcolor}
\usepackage{physics}
\usepackage{tikz}
\usetikzlibrary{tikzmark}
\usepackage{dashbox}%
\usepackage{mathtools, amssymb}
\usepackage[colorlinks=true]{hyperref}
\usepackage{orcidlink}
\usepackage{changes}
\usepackage{booktabs}

\newcounter{cycle}

\begin{document}
\title{Free Snacks in Quantum Complexity}

\begin{abstract}
Estimating ground-state energies is a cornerstone problem in Hamiltonian complexity, and in general requires exponential resources even on quantum computers. It is in this context we analyse the recently developed Imaginary-Time Quantum Dynamical Emulation (ITQDE)\cite{leamer_quantum_2024}. This method enables estimation of spectral densities, partition functions, and low-lying gaps, but requires only minimal coherent control, modest classical post-processing, and no state preparation. Using a quadrature-based formulation, we derive scaling and stability criteria that diagnose when its estimates are reliable, and introduce a controlled smoothing that yields a principled bias–variance trade-off. The resulting picture preserves the hardness of exact eigenvalue resolution but reveals a practical regime - a “free snack” -  where coarse-grained spectral information is obtainable with only polynomial resources. By recasting sampling costs as explicit bounds on resolvable bandwidths, the intermediate regime between trivial and intractable complexity becomes accessible on near-term quantum hardware.
\end{abstract}

\author{Gerard McCaul\orcidlink{0000-0001-7972-456X}}
\affiliation{Department of Physics, Loughborough University, Loughborough, UK}
\email{g.mccaul@lboro.ac.uk}


\maketitle

\section{Introduction}
All choices carry costs. This is commonly expressed via the dismal science's adage that “there is no such thing as a free lunch” \cite{friedman1975:free_lunch,ade2016:free_lunch}, but is equally true in computation as in finance. There is no universal algorithmic solvent, and any performance gain must be purchased at the expense of some other resource \cite{wolpert_no_1997}.  In classical computer science this trade-off is crystallised in the hardness of NP-complete problems, where the cost of all known algorithmic solutions scale exponentially with problem size.  

Quantum computing is often framed as a solution to these constraints, in which computationally hard problems can be efficiently solved. The reality of this is more subtle, and in many instances remains in dispute. An illustrative example is computation via quantum optical interferometry. This not only enables integer factoring  \cite{clauser_factoring_1996}, but - in principle - permits solutions to NP-complete tasks to be computed in polynomial time \cite{shaked_optical_2007,cerny_quantum_1993}. The inevitable catch is that the number of photons required - and therefore the energetic cost - grows \textit{exponentially} with problem size. More broadly, many NP problems can be directly realised by Ising models, such that their ground state encodes the solution \cite{lucas_ising_2014}. Unfortunately, ground-state estimation remains a `hard' problem \cite{cubitt_complexity_2016, kempe_complexity_2006, gharibian_quantum_2015, osborne_hamiltonian_2012} for quantum computers. Even with quantum resources, there is no free lunch \cite{volkoff_universal_2021}. 

One arena in which quantum computers have provable utility is in \textit{Hamiltonian simulation} \cite{lloyd1996universal, zalka_simulating_1998}. While in its full generality the problem remains hard \cite{osborne_hamiltonian_2012}, the dynamics of \textit{local} Hamiltonians may be efficiently simulated on a quantum circuit. This immediately raises the question of whether the spectral information of a given Hamiltonian can be extracted with access to efficient dynamical simulation. Adiabatic quantum computing \cite{aharonov2007adiabatic,farhi2001quantum} approaches this by slowly evolving an easy initial Hamiltonian in its ground state to a target Hamiltonian encoding the task. The adiabatic theorem ensures success provided the interpolation is slow relative to the inverse square of the minimum spectral gap. If this is polynomially bounded then the ground-state can be prepared efficiently. Typically however the minimal gap closes exponentially, restoring the complexity barrier associated with generic ground-state estimation \cite{kempe_complexity_2006}. 

An alternative approach to extracting spectral information is through \textit{imaginary-time evolution} \cite{mcmahon_equating_2025}. This is a form of \textit{non-unitary} dynamics, which cannot be directly implemented with unitary gates. For this reason approaches to simulating non-unitary dynamics in a universal quantum computer have been a subject of intense interest  \cite{HamSim, Schrodingerisation, an_linear_2023, NOVIKAU2025109498}. In the case of imaginary-time simulation, its promise lies in the fact that spectral information (including ground states, spectral gaps and partition functions) may be calculated directly from imaginary dynamics. While a number of approaches to spectral calculation have been developed - a non-exhaustive list includes quantum phase estimation \cite{parker_quantum_2020, kitaev_classical_2002, nielsen_quantum_2010, svore_faster_2013, obrien_quantum_2019, wiebe_efficient_2016, dobsicek_arbitrary_2007}, the rodeo algorithm \cite{choi_rodeo_2021, cohen_optimizing_2023}, Fourier-transform-based methods \cite{somma_quantum_2020,gnatenko_detection_2022, gnatenko_energy_2022}, and variational \cite{xie_variational_2024, jones_variational_2019} and feedback-based \cite{rahman2024feedback} methods - imaginary dynamics remain especially appealing because of the breadth of quantities that can be directly inferred from it. It is however this very quality which implies exponential costs in the implementation of any imaginary-time based method. 

It is against this backdrop that we investigate a recently developed technique, termed Imaginary-Time Quantum Dynamical Emulation (ITQDE) \cite{leamer_quantum_2024}. ITQDE exploits a correspondence that maps non-unitary filters onto an ensemble of unitary evolutions. Superficially this resembles the Linear Combination of Unitaries (LCU) framework \cite{an_linear_2023}, but ITQDE and LCU have fundamentally distinct requirements. LCU requires coherent control via ancilla qubits and amplitude amplification, and these steep overheads make it more suited to future fault-tolerant devices. In contrast to this ITQDE is a \textit{constructive} mapping, providing a deterministic, sampling-based route to spectral information. It requires no state preparation, minimal coherent control, and only a modest degree of classical post-processing. More strikingly, this not only permits the calculation of thermal states, but grants access to the full Hamiltonian spectrum from a single set of measurements. The apparent economy with which ITQDE can extract this information is puzzling, and naturally invites skepticism. The central purpose of this work is therefore to articulate the precise capacities and limitations of ITQDE, to delineate the regimes in which it is reliable, and to provide explicit guarantees on accuracy, stability, and resource scaling. 

TO achieve this, we first recapitulate in Sec. \ref{sec:spectral_staircase} how ITQDE can be implemented as a quantum algorithm, from which a “spectral staircase” estimator for Hamiltonian eigenvalues can be extracted. This is followed in Sec. \ref{sec:complexity} by a discussion of the computational complexity inherent in spectral calculation, and a survey of the hidden exponential costs that reside in all imaginary-time based algorithms. By identifying the origin of these costs in ITQDE, we develop in Sec. \ref{sec:quadrature} a quadrature–based approximation that radically reduces the implementation costs of ITQDE. The underlying stability of this implementation is then investigated in Sec. \ref{sec:quadrature_stab}, and then employed in Sec. \ref{sec:sampling} to derive the conditions under which sampling costs become exponential. Finally, we show that post-processing can be used to formulate a principled trade-off, where coarse-grained spectral information can be extracted with modest resources. 

The result is a picture in which the computational complexity of exact ground-state estimation remains inviolate, but nonetheless admits loopholes in the form of approximate, resource-efficient probes. Rather than a free lunch, ITQDE offers a “free snack”: a tractable route to spectral and thermodynamic data on near-term devices, opening a middle terrain between trivial simulation and intractable complexity. 

\section{The Spectral Staircase \label{sec:spectral_staircase}} 
We begin with a recapitulation of the recently developed Imaginary-Time Quantum Dynamical Emulation (ITQDE) correspondence \cite{leamer_quantum_2024}. This reformulates an imaginary-time evolution as a sum of experimentally accessible overlaps of oppositely propagated states. Put differently, it is a constructive equivalence between the expectations of a non-unitarily propagated state and an ensemble of unitary evolutions. The consequence of this is that ITQDE enables a sampling-based approach to the calculation of spectral properties in quantum circuits. 

To understand how this can be achieved, we first consider a system described by a Hamiltonian $H$, with an initial state $\ket{\psi_0}$. We then define the following: 
\begin{equation}
\label{eq:specstaircase}
   H_\tau(\lambda)   =\frac{N_\tau (\lambda)}{Z_\tau (\lambda)}= \frac{\Braket{\psi_0|H e^{-\tau(H - \lambda)^2}|\psi_0}}{\Braket{\psi_0|e^{-\tau(H - \lambda)^2}|\psi_0}} 
\end{equation}

This quantity has a number of interpretations, but in the present context it is best understood as a form of \textit{spectral filtering} around a probe-point $\lambda$ by a Gaussian of width $\frac{1}{\sqrt{\tau}}$. Let us say $H$ has a set of spectral levels $E_j$ with degeneracy $g(E_j)$, and presume the state $\ket{\psi_0}$ has nonzero overlap with each eigenstate (i.e., is not itself an energy eigenstate). Writing the initial populations as \(p_j=|\langle E_j|\psi_0\rangle|^{2}\), we find with slight rearrangement that Eq.~\eqref{eq:specstaircase} is simply a \emph{weighted mean}
\begin{equation}
 H_\tau(\lambda)= \sum_{k}E_k P_k(\lambda,\tau)
  \label{eq:weighted_mean}
\end{equation}
with respect to the distribution 
\begin{equation}
 P_j(\lambda,\tau)= \frac{p_j\,g(E_j)\,
  e^{-\tau\,(E_j-\lambda)^{2}}}{\sum_k p_k\,g(E_k)\,
  e^{-\tau\,(E_k-\lambda)^{2}}}.   
\end{equation}

Now consider two successive levels $E_j<E_{j+1}$, separated by a central gap \(\Delta_j=\frac{1}{2}(E_{j+1}-E_j)\) and midpoint \(\bar E_j=\tfrac12(E_j+E_{j+1})\). In the region \(|\lambda-\bar E_j|<\Delta_j\), Eq.\eqref{eq:weighted_mean} may be approximated with exponential accuracy with only the truncated sum across levels \(k=j,j\!+\!1\). In this span of $\lambda$ we have a \textit{logistic} curve
\begin{equation}
  H_\tau (\lambda) \simeq
  E_j + 
  2\frac{\Delta_j}{
        1 +
        R_j\,e^{\tau\Delta_j(\lambda-\bar{E}_j)}}, 
  \label{eq:logistic}
\end{equation}
where $R_j=\frac{p_j\,g(E_j)}{p_{j+1}g(E_{j+1})}$. This is a nascent step function between the levels $E_j$ and $E_{j+1}$. The width of the transition between these plateaux will be
\(\delta\lambda\sim(\tau\Delta_j)^{-1}\), meaning
for \(\tau\Delta_j\gg1\) the crossover becomes sharp. In the asymptotic limit $\tau\to\infty$, $H_\tau(\lambda)$ approaches a piecewise constant function of $\lambda$, tracing out a \emph{spectral staircase}. That is, as $\lambda$ is varied, $H_\tau(\lambda)$ steps through plateaux located at $E_j$, with the location of each inflection point between levels depending on the relative state populations and degeneracies.

For finite but large~$\tau$, the estimator thus encodes spectral information smoothed over a scale $\sim\tau^{-1/2}$, and we obtain a \textit{spectral staircase}. That is, as $\lambda$ is varied $ \langle H_\tau^{(\lambda)}\rangle$ steps through plateaux located at the spectral levels $E_j$, while the sharp inflection point between levels depends on the initial state populations and degeneracies. We can bound the accuracy of any eigenenergy inferred from this staircase by considering the bound in a region $|\lambda-E_j|<\Delta_j$:
\begin{equation}
  \bigl|H_\tau (\lambda)- E_j\bigr|
  \;\le\;
  \Delta_j\,e^{-\tau \Delta_j^{2}}.
  \label{eq:uniform_bound}
\end{equation}
Rearranging this inequality, we find that resolving $E_j$ to accuracy $\varepsilon$ from $H_\tau(\lambda)$ requires $\tau\geq
  \frac{\ln(\Delta_j/\varepsilon)}{\Delta_j^2}$. We may therefore estimate that resolving a given level from its neighbours requires $\tau\gtrsim1/\Delta_j^{2}$. 

From this analysis, it is clear that the variation of the spectral staircase $H_\tau (\lambda)$ with $\lambda$ contains physically pertinent features of $H$, including both its spectrum and (via $R_j$) its density of states, with the resolution of this information controlled by $\tau$.

\subsection*{Building the Staircase}
This begs an obvious question — how might one infer this spectral staircase in a measurement or quantum-algorithmic context? The practical challenge is that the filter $e^{-\tau(H-\lambda)^2}$ is in effect an imaginary-time propagation under $H^2$. Such a manifestly \emph{non-unitary} evolution that cannot be executed directly on coherent quantum hardware or — more broadly — any experiment constrained to unitary dynamics. The core contribution of ITQDE is to bypass this obstacle by recasting the non-unitary propagator as an overlap of two \emph{unitary} evolutions. While the full details of the ITQDE correspondence are left to Ref. \cite{leamer_quantum_2024}, it will suffice here to quote the result. Discretising $\tau=m\Delta\tau$, we then define a unitary operator
\begin{equation}
	\mathcal{U} = e^{-i\sqrt{\frac{\Delta\tau}{2}}H},
 \label{eq:U}
\end{equation}
and introduce the following notation for its application to states (with initial state $\ket{\psi_0}$):
\begin{align}
	|\psi_{j+1}\rangle &= {\mathcal{U}}|\psi_j\rangle,  & |\psi_{j-1}\rangle &= {\mathcal{U}}^\dag|\psi_j\rangle.
\end{align}
Note that the explicit role of both forward and backward evolutions in ITQDE echoes the bidirectional evolution central to DB-QITE, and is redolent of recent algorithms exploiting time symmetry for eigenstate determination \cite{wei_time-symmetric_2025}.

With these recursive definitions of oppositely evolved states, we 
are now in a position to state the ITQDE correspondence:
\begin{equation}
H_\tau(\lambda)  \approx \langle H_\tau^{(\lambda)}\rangle =  \frac{\langle N^{(\lambda)}_\tau\rangle} {\langle Z^{(\lambda)}_\tau \rangle}
\end{equation}
where the functions $N^{(\lambda)}_m$ and $Z^{(\lambda)}_m$ are \textit{discrete approximations} to the numerator and denominator of Eq.\eqref{eq:specstaircase}. Each of these terms is approximated to $\mathcal{O}(m\Delta\tau^2)$ by 
\begin{equation}
 \langle N^{(\lambda)}_\tau \rangle=\frac{1}{2^m}\sum_{j=0}^{\frac{m}{2}}e^{2i\lambda j\sqrt{\frac{\Delta\tau}{2}}}{{m}\choose {\frac{m}{2}+j}}\langle\psi_{2j}|H|\psi_{-2j}\rangle + \text{c.c.},
  \label{eq:shifted_N}
 \end{equation}
 and
\begin{equation}
	\langle Z^{(\lambda)}_\tau \rangle= \frac{1}{2^m}\sum_{j=0}^{\frac{m}{2}}e^{2i\lambda j\sqrt{\frac{\Delta\tau}{2}}}{{m}\choose {\frac{m}{2}+j}}\langle\psi_{2j}|\psi_{-2j}\rangle + \text{c.c.} \label{eq:shifted_z}
\end{equation}

\begin{figure}
    \centering
    \includegraphics[width=1.1\linewidth]{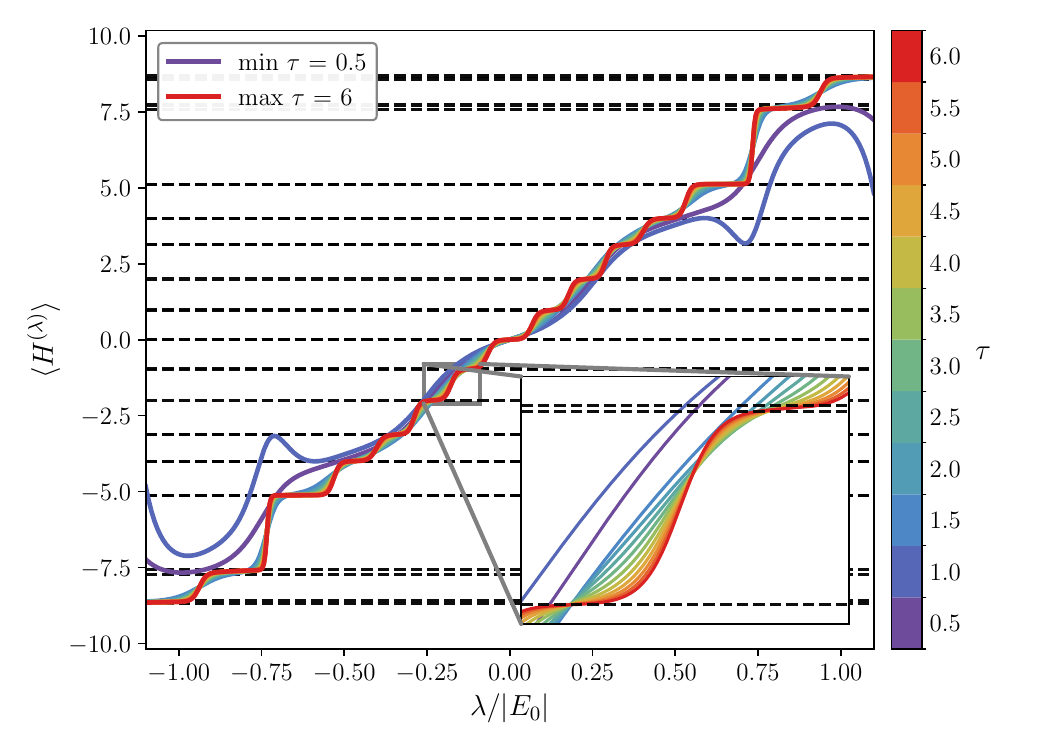}
    \caption{Plot of the filtered expectation value $\langle H_\tau ^{(\lambda)}\rangle$ (calculated via ITQDE) for a 2D-FH model where $U=2$,  $L_x=2$, $L_y =2$ and half-filling. As $\tau$ is increased the staircase sharpens into near-vertical steps (red), converging toward plateaux at the true eigen-energies (black dashed lines). Inset: zoom on one transition, illustrating the transitory width contraction with growing $\tau$.}
    \label{fig:staircase_example}
\end{figure}

 The significance of this correspondence lies in the fact that it is a \textit{constructive} weighting of overlaps of the form $\langle\psi_{2j}|O|\psi_{-2j}\rangle$.  A number of approaches exist for the calculation of such quantities, ranging from the standard Hadamard test \cite{nielsen2010quantum} to more NISQ-friendly interferometric or shadow-based estimators \cite{elben_cross-platform_2020,huang_predicting_2020}, such as randomized-measurement overlaps, direct fidelity estimation \cite{flammia_direct_2011}, or echo-style Loschmidt protocols \cite{garttner_measuring_2017}. These are all implementable with shallow circuits that do not require controlled gates. Then, because each overlap $\langle\psi_{2j}|O|\psi_{-2j}\rangle$ is \emph{independent of $\lambda$}, a \emph{single} set of measurements suffices. Once the collection of overlaps has been estimated, then Eq.\eqref{eq:shifted_N} and Eq.\eqref{eq:shifted_z} can be evaluated for \textit{arbitrary} probe values $\lambda$ by classical re-weighting. Scanning $\lambda$ across the energy range therefore generates the spectral staircase. In this manner, a set of experimentally accessible overlaps can be used to characterise the spectrum of a quantum system, trading the preparation of many distinct quantum circuits for inexpensive classical post-processing.

We conclude this section with a numerical illustration of the spectral staircase. In this and all other cases we use the 2D Fermi-Hubbard model \cite{imada_1998, anderson_1987,hofrichter_2016}. Its Hamiltonian is given by
\begin{equation}
    H = \sum_{\langle i,j\rangle,\sigma}t_{ij}(c_{i\sigma}^\dag c_{j\sigma} + \text{h.c.}) + U\sum_{j=0}^{L-1} n_{j\uparrow}n_{j\downarrow} - \mu \sum_{j=0}^{L-1} n_{j\sigma},
\end{equation}
where $t_{ij}$ denotes the tunneling amplitude between lattice sites $i$ and $j$, $c_{j\sigma}^\dag$ and  $c_{j\sigma}$ are the creation and annihilation operators associated with a fermion of spin $\sigma$ at lattice site $j$, respectively, $U$ characterizes the on-site interaction of fermions, $n_{j\sigma}=c^\dag_{j\sigma}c_{j\sigma}$ is the occupation number operator, $\mu$ is the chemical potential, and $L=L_xL_y$ is the number of lattice sites.   

Applying Eqs.(\ref{eq:shifted_N},\ref{eq:shifted_z}) to this model, its spectral staircase can be inferred from $\langle H_\tau^{(\lambda)}\rangle$. Examples of this are shown in Fig.\ref{fig:error_curves}, where increasing $\tau$ sharpens the steps of the staircase.

\section{Complexity in Spectral Estimation \label{sec:complexity}}
We have established that Eq.\eqref{eq:specstaircase} can be employed to infer spectral information, and that ITQDE implements this as a quantum algorithm \cite{leamer_quantum_2024}. On the face of it, this appears to offer an improbably efficient route to spectral information. ITQDE is a deterministic method that requires no state preparation or variational optimisation. Moreover, for any $k$-local observable $O$, each overlap $\langle\psi_{2j}|O|\psi_{-2j}\rangle$  can be efficiently implemented as a set of quantum circuits whose depth scales polynomially with system size. If this were the whole story, it would amount to a loophole in one of the most entrenched results in quantum complexity theory: the hardness of the local Hamiltonian problem \cite{kitaev_classical_2002,kempe_complexity_2006, gharibian_quantum_2015, osborne_hamiltonian_2012}.

This decision problem asks whether the ground-state energy of an $n$-qubit $k$-local Hamiltonian lies below or above a given threshold. For $k\ge2$ this problem is QMA-complete (the quantum analogue of NP-completeness) and therefore widely believed to be intractable for polynomial-time quantum algorithms in the worst case.  The operational implication of this is that any algorithm claiming to resolve $E_0$ to additive error $\varepsilon = \operatorname{poly}(1/n)$  will require resources that scale \emph{exponentially} with~$n$.

There is ample evidence of this barrier when surveying the landscape of imaginary-time techniques developed in recent years. Variational quantum algorithms have been developed, where a parameterized quantum circuit is optimized to give the best approximation to an imaginary-time evolution \cite{gomes_adaptive_2021, gomes_efficient_2020, mcardle_variational_2019}. These variational approaches can result in shallow quantum circuits that are favorable for near-term hardware implementations, but the computational costs of the classical optimization rapidly become intractable for larger systems due to the existence of local minima and barren plateaus. 

Another leading example is the QITE algorithm \cite{motta_determining_2020, kamakari_digital_2022, tsuchimochi_improved_2023, sun_quantum_2021}, where an evolution in imaginary time is approximated using a Trotter decomposition of a real-time evolution, i.e., where the latter is expressed as the product of unitary, real-time evolutions over a sequence of small time steps. The generators of these real-time evolutions are determined sequentially, based on tomographic measurements performed at each step. This measurement cost can become prohibitive for systems that develop long-range correlations. A recent extension of QITE replaces its variational fit to a local unitary with a shallow double-bracket circuit \cite{gluza_double-bracket_2025}. While this guarantees a monotonic, iterative approach to the ground state, it also incurs an exponential cost in circuit depth. 

A third approach is the PITE algorithm \cite{PhysRevResearch.4.033121,kosugi_exhaustive_2023, turro_imaginary-time_2022, nishi_optimal_2023, xie_probabilistic_2022}, where the imaginary time evolution is encoded into a unitary operation acting on a larger Hilbert space. After evolving the total system one step using this larger unitary operator, an ancilla qubit can be measured to determine whether the step of imaginary time evolution was successfully applied or not. If so, the procedure can be repeated to evolve over the next time step, and if not, it must be restarted.  Thus, evolving to later times results in an exponential decrease in the probability of successfully evolving the original system in imaginary time. Some recent works \cite{liu_probabilistic_2021, nishi_acceleration_2022} address this situation by applying amplitude amplification after each step, but the associated costs can become impractical as the number of steps increases.

This complexity barrier, and the exponential scaling it manifests is highlighted in Table \ref{tab:complexity_walls}, which sketches how bottlenecks arise in an exemplar selection of imaginary-time algorithms. Like ITQDE, these can be applied to the calculation of ground-state energies, and in each case a potential exponential cost is readily identified.    

\begin{table*}[hbt!]
\caption{Representative algorithms for imaginary–time based algorithms for ground state preparation and energy estimation. Their mechanism of action and the origin of their worst–case exponential cost are briefly described.}
\label{tab:complexity_walls}

\begin{ruledtabular}
\renewcommand{\arraystretch}{1.15}  
\begin{tabular}{p{2.8cm}p{5.1cm}p{5.1cm}}
\textbf{Algorithm} & \textbf{Mechanism} & \textbf{Complexity (worst case)} \\ \hline
QITE\,\cite{gomes_adaptive_2021,gomes_efficient_2020,mcardle_variational_2019,motta_determining_2020} &
Approximate each small imaginary-time step $e^{-\Delta\tau H}$ by fitting a \emph{local} unitary acting on a cluster of size $L\!\sim\!\xi$ (correlation length). Requires solving a linear system of expectation values at every step. &
Number of observables and shots grows as $d^{L}$; for critical or highly-entangled Hamiltonians $\xi\!\sim\!O(n)$, giving total measurements $\mathcal{O}(\exp(n))$.  Classical post-processing of the linear system scales identically, so runtime is exponential in $n$ when $\xi$ scales. \\[4pt]
DB-QITE\,\cite{gluza_double-bracket_2025} &
Implements the double-bracket flow $d\rho/dt=[\rho,[\rho,H]]$ by compiling \emph{each} $\Delta\tau$ step into a shallow circuit; fidelity increases monotonically. &
Energy variance decreases $\propto 1/k$ after $k$ iterations, so reaching accuracy $\varepsilon$ needs $k\sim\mathcal{O}(1/\varepsilon)$.  Compiled circuit depth \emph{doubles} every iteration. Implies a depth of $\mathcal{O}(\exp(1/\varepsilon))$ for high precision. Likewise ancilla/qubit count if unitaries are laid out serially. \\[4pt]
PITE\,\cite{PhysRevResearch.4.033121,kosugi_exhaustive_2023, turro_imaginary-time_2022, nishi_optimal_2023, xie_probabilistic_2022} &
Realises a non-unitary step by post-selecting an ancilla after controlled $e^{\pm iH\Delta t}$ evolutions (“repeat-until-success’’).  Chain of many successful steps approximates $e^{-\tau H}$. &
Per-step success $\simeq\!e^{-\Delta t\,\Delta_0^2}$ ($\Delta_0=E_1-E_0$).  To reach total $\tau=1/\varepsilon$ needs $\tau/\Delta t$ steps; expected repetitions $\exp(\tau\Delta^{2})$. Worst-case $\Delta_0 \sim2^{-n}$ implies $\exp(2^{n})$ repeats required.  Amplitude-amplification variants cut \emph{polynomials} only. \\[4pt]
RODEO\,\cite{choi_rodeo_2021,cohen_optimizing_2023} &
Sequentially rotates and measures an ancilla; each round multiplies off-target amplitudes by $\cos(\Delta E\,t)$, exponentially amplifying a chosen eigenstate whose energy is within $\varepsilon$ of a probe value $E$. &
Rounds $r\!\sim\!O[(\log1/\varepsilon)/(\varepsilon t)]$, with $t\!\sim\!1/\varepsilon$ for resolution $\varepsilon$.  Runtime $\propto(\log1/\varepsilon)^{2}/(p\,\varepsilon)$.  If initial and ground state overlap $p\!\sim\!2^{-N}$ or one demands $\varepsilon\!\sim\!2^{-\mathrm{poly}(N)}$, cost blows up to $\exp(n)$ or worse.  \\ 
\end{tabular}
\end{ruledtabular}
\end{table*}

This scaling behaviour will be generic to any quantum algorithm that is capable of ground-state energy estimation, meaning there is no spectral stairway to heaven, and ITQDE must hide an exponential cost. For it to be of practical use, it is necessary to understand both where its bottlenecks lurk and how they might be negotiated.  

Identifying the source of this scaling is not without its challenges however. This is because the ITQDE correspondence is philosophically and practically distinct from the methodologies outlined in Table \ref{tab:complexity_walls}. It is a constructive correspondence, and its only \textit{algorithmic} component stems from the ensemble of independent circuits used to calculate each overlap. Moreover, it employs several free parameters (i.e. $\tau$, $\Delta \tau$) which control both the resolved energy scale and discretisation error. The scaling for the number and depth of circuits required to calculate an individual overlap $\langle\psi_{2j}|O|\psi_{-2j}\rangle$ is only polynomial, meaning there are only two possible sources of exponential scaling. These are the number of overlaps $m$, and the sampling budget for estimating each overlap. These ultimately control the accuracy to which ITQDE approximates $H_\tau (\lambda)$, and therefore must contain the mechanism through which the complexity wall manifests itself.  

As a preliminary, it is necessary to outline the generic dependence of spectral properties on system size. A $k$-local Hamiltonian will decompose into $H=\sum_{j=1}^{r}h_{j}$ terms, with $r\le\binom{n}{k}=O(n^{k})$. By definition the local terms have norm $\|h_j\|\le \Lambda=\mathrm{poly}(n)$ \cite{kempe_complexity_2006, kitaev_classical_2002}, and hence
$\|H\|\le r\Lambda = O(\Lambda\,n^{k})$.
For physically motivated models $\Lambda = O(1)$, and the spectral norm will scale as $\|H\| = O(n^{k})$. The corresponding Hilbert space dimension $D$ will go as $D\sim \mathcal{O}(2^n)$. For a generic non-integrable system, one expects level statistics similar to random matrix ensembles, resulting in a dense unstructured spectrum \cite{haake2010,mehta2004}. We therefore expect that the typical energy gap shrinks exponentially with $n$ i.e. $\Delta_j \sim  \mathcal{O}(\frac{n^k}{2^n})$. 

The consequence of this for ITQDE can be inferred from the fact that resolving energies separated by a gap $\Delta_j$ requires $\tau \gtrsim 1/\Delta^2_j$. The exponential shrinking of this gap with $n$ therefore demands a reciprocal scaling in the $\tau$ necessary to fully resolve the spectrum. If — as outlined in Sec.\ref{sec:spectral_staircase} and Ref.\cite{leamer_quantum_2024} — the discretisation accuracy is bounded by $\mathcal{O}(\tau \Delta \tau)$, the number of overlaps $m=\frac{\tau}{\Delta \tau}$ required to keep this constant will scale quadratically with respect to $\tau$. This implies that for a generic system, the number of overlaps needed to fully resolve the spectrum must scale exponentially with system size. 

Note however that this conclusion follows directly from the densification of the spectrum with $n$, and the corresponding scaling of $\tau$ required to resolve individual levels. It says nothing about the cost of the method if one instead chooses a \textit{fixed} resolution. Moreover, the Gaussian nature of the filter lends itself to a number of highly accurate and inexpensive approximation schemes. These can be used to greatly reduce the number of overlaps required by ITQDE, and the scaling of its implementation cost.  

\section{Quadrature Approximated ITQDE\label{sec:quadrature}}
We have seen that fully resolving the spectral staircase requires an exponential number of overlaps to fully resolve the spectrum, and the exponential compression of typical gap sizes. Remarkably however, it is possible to reduce the number of overlaps required such that they scale \textit{linearly} with $n$. To do so, we first appeal to the fact (established in Ref.\cite{leamer_quantum_2024})  that the continuous $m\to \infty$ limit of ITQDE recovers a form of \textit{Hubbard-Stratonovich} transformation. We may take this limiting integral relationship

\begin{equation}
e^{-\tau H^2} = \frac{1}{\sqrt{\pi}} \int_{-\infty}^\infty e^{-x^2} e^{-2i\sqrt{\tau}x H} dx.
\end{equation}
and replace it with a Gauss-Hermite quadrature approximation \cite{NIST:DLMF}. This permits the definition of a \textit{quadrature approximated} ITQDE correspondence:
\begin{align}
\label{eq:ITQDE-Gauss}
\langle H_m^{(\lambda)}\rangle &= \frac{\langle N_m^{(\lambda)}\rangle}{\langle Z_m^{(\lambda)} \rangle} \\
\label{eq:N-Gauss}
 \langle N_m^{(\lambda)} \rangle &= \sum_{k=1}^m w_k \mathrm{Re}[{\rm e}^{2i \lambda \tau_k} \langle \psi_{-\tau_k} | H | \psi_{\tau_k} \rangle], \\
 \label{eq:Z-Gauss}
\langle Z_m^{(\lambda)}\rangle &= \sum_{k=1}^m w_k \mathrm{Re}[{\rm e}^{2i \lambda \tau_k} \langle \psi_{-\tau_k} | \psi_{\tau_k} \rangle]
\end{align}
 Here we commit a slight abuse of notation, and distinguish quadrature approximated quantities by index $m$ subscripts rather than the $\tau$ employed in Eqs.(\ref{eq:shifted_N},\ref{eq:shifted_z}). The quadrature approximation employs a set of $m$ points $x_k$, with associated $w_k$ weights. The new time-points for the overlaps are then set by $\tau_k=\sqrt{\tau}x_k$, and the evolved state is notated as:
 \begin{equation}
     \ket{\psi_{\pm \tau_k}}= e^{\mp i\tau_k H}\ket{\psi_0}
 \end{equation}

App. \ref{app:quad_approx} establishes the asymptotic error of this quadrature discretisation is $ \mathcal{O}(\frac{\tau \| H \|^{2}}{m})^m$. Critically this means that the quadrature approximates the Gaussian filter exponentially accurately in $m$. Moreover, the weights $w_k$ associate with a degree $m$ Gauss-Hermite polynomial, and will themselves be exponential. This implies that only a fraction $\bar{m} < m$ of terms meaningfully contribute. The error $\varepsilon$ that results from pruning the tail of the quadrature will be:
\begin{equation}
\label{eq:quad-error}
   \varepsilon= \sum_{k=\bar{m}+1}^m w_k f_k
\end{equation}
where $f_k$ stands in for any overlap based terms in the summand. For $k\gg 1$, the $k$-th positive Hermite node and its weight satisfy  
$x_k \simeq \sqrt{2k}$  and  $w_k \simeq C\,e^{-x_k^{2}}\simeq C\,e^{-2k}$ \cite{NIST:DLMF}. The truncation error therefore obeys $\varepsilon \sim e^{-d\bar m}$, where $d\approx 2$. This exponential dependence immediately suggests that one can set the quadrature degree $m$ arbitrarily highly, but employ the $\bar{m}$ most heavily weighted overlaps in Eqs.(\ref{eq:N-Gauss}-\ref{eq:Z-Gauss}). We denote these truncated quantities by the reduced $\bar{m}$, e.g. $\langle H_{\bar{m}}^{(\lambda)}\rangle$. 

We can infer how the number of truncated overlaps required must scale as follows. First, we define the factor $s:=\tau\lVert H\rVert^{2}$, such that the asymptotic quadrature error is 
$\varepsilon_{\rm{quad}}\sim (s/m)^{m}$. Matching this to the truncation $\varepsilon_{\rm quad} \sim \varepsilon$ requires 
\begin{equation}
(s/m)^{m} \;\sim\; e^{-d\,\bar m}.
\end{equation}
Choosing $m=\kappa s$ with fixed $\kappa>1$ gives 
$(s/m)^{m} = \kappa^{-\kappa s} = e^{-\gamma s}$, 
where $\gamma=\kappa\ln\kappa>0$.  
Balancing exponents yields $\frac{\bar m}{s}  \sim \frac{\gamma}{d}$.
Equivalently, for fixed $\kappa$, we have $\varepsilon^{1/s} \;\approx\; e^{-\,d\,\bar m/s}$. From this we can infer a scaling for $\bar{m}$ with $s$ necessary for a fixed precision $\varepsilon$: 
\begin{equation}
    \bar{m} \propto s \log(\varepsilon^{\frac{1}{s}}).
\end{equation}

\begin{figure}
    \centering
    \includegraphics[width=\linewidth]{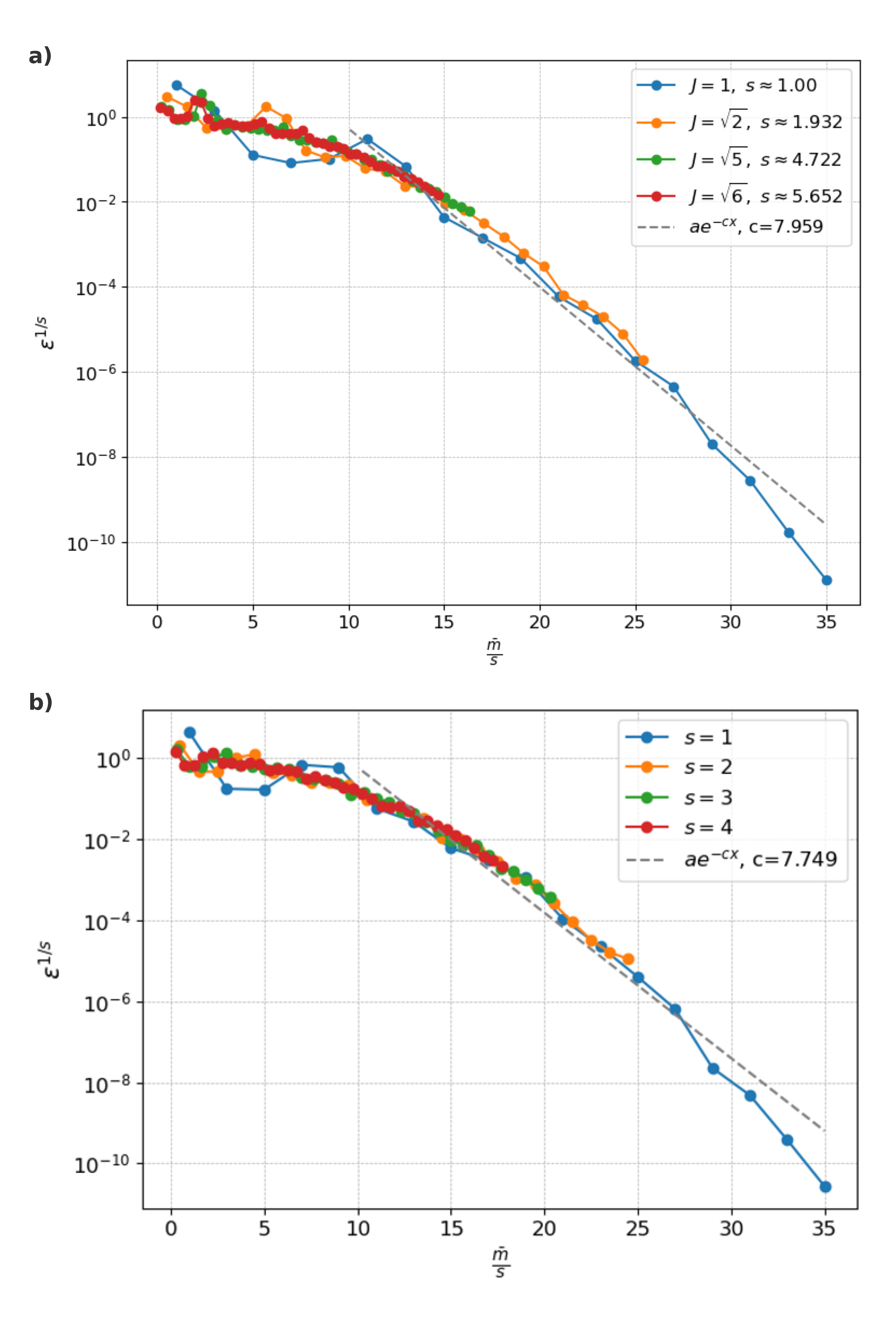}
    \caption{Error scaling of the ITQDE correspondence calculated via Gaussian quadrature using $\bar{m}$ overlaps, applied to the 2D-FH model using $U=2$,  $L_x=2$, $L_y =3$ and half-filling. \textbf{(a)} Scaling of quadrature error as the spectral norm $s = \|H\|^2$ is increased via hopping strength $J$. The scaled error $\epsilon^{1/s}$ is plotted against the rescaled quadrature depth $\bar{m}/s$, confirming exponential suppression with linear overlap growth.  
\textbf{(b)} Equivalent scaling collapse as $\tau \to s \tau$ is varied for a fixed Hamiltonian. In both cases, the empirical scaling matches the theoretical prediction $\epsilon^{1/s} \sim e^{-c\,\bar{m}/s}$ stemming from Gauss-Hermite truncation.
}
    \label{fig:error_curves}
\end{figure}

To test this proposition, we denote Eq.\eqref{eq:ITQDE-Gauss} truncated to $\bar{m}$ as $\langle H_{\bar{m}}^{(\lambda)}\rangle$, and define an integrated error between the true spectral staircase  and its quadrature approximation:
\begin{equation}
\label{eq:error}
    \epsilon= \int^{E_{\rm max}}_{E_{\rm min}} {\rm d}\lambda \left|\langle H_{\bar{m}}^{(\lambda)}\rangle
    -   H_{\tau}(\lambda)\right|.
\end{equation}
Fig.\ref{fig:error_curves} plots this, demonstrating the predicted scaling relationship predicted between $\bar{m}$ and $s$ collapses the error curves. This logarithmic scaling of $\bar{m}$ serves to highlight the surprising parsimony with which ITQDE may be implemented when employing quadrature. Comparing Fig.\ref{fig:error_curves} to analogous linear calculations in Ref.\cite{leamer_quantum_2024}, the error across the spectrum is reduced by many orders of magnitude while simultaneously requiring $\sim 2$ orders of magnitude fewer overlaps.  Most significantly, this means that while $\tau$ must scale exponentially with $n$, the effective number of overlaps required to approximate the staircase grows only linearly with $\tau$, meaning $\bar{m}\sim \mathcal{O}(n)$. 

\section{Quadrature Stability and Spectral Resolution \label{sec:quadrature_stab}}
We now examine the question of whether truncation might introduce numerical instabilities into the calculation of the spectral staircase. There are a number of mechanisms by which this can occur, but in all cases they depend strongly on the behaviour of the truncated partition function. To investigate this, it is convenient to suppress dependence on the initial state $\ket{\psi_0}$ by replacing any expectation with respect to it by a trace. This is done solely to simplify the analysis, and modifies (for example) the truncated approximation of the partition function to:
\begin{equation}
\label{eq:Z_trace}
\langle Z_{\bar{m}}^{(\lambda)} \rangle \approx \sum_{k=1}^{\bar{m}} w_k\, \mathrm{Tr}\left[ e^{-i H \tau_k} \right]\, e^{2i \lambda \tau_k}
= \sum_{j=1}^{D} \sum_{k=1}^{\bar{m}} w_k\, e^{-i(E_j - \lambda) \tau_k}.
\end{equation}

\begin{figure}[h]
    \centering
\includegraphics[width=\linewidth]{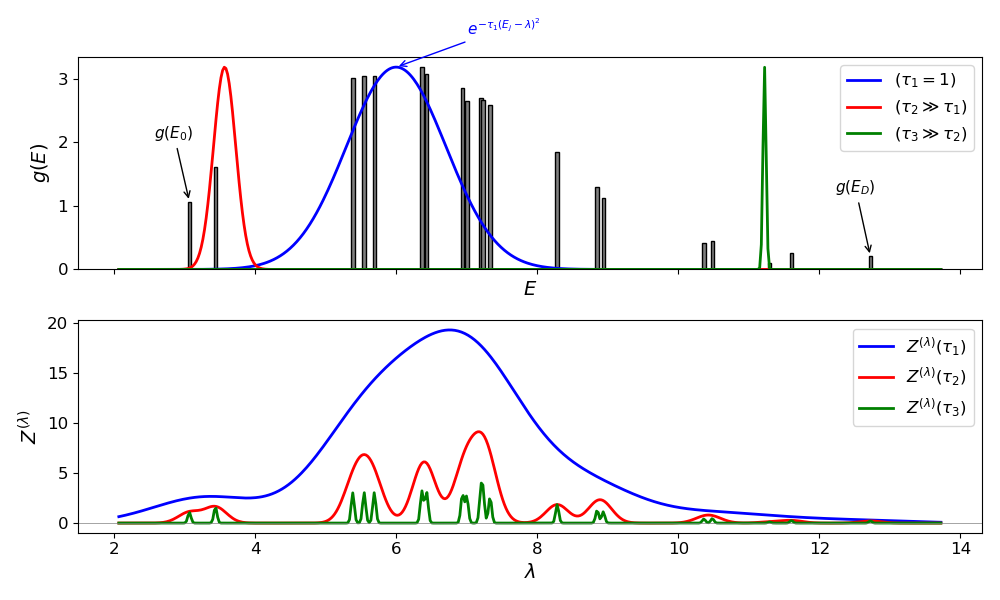}
    \caption{\textbf{Top:} Illustration of spectral filtering via the Gaussian kernel $e^{-\tau(E_j - \lambda)^2}$ applied at increasing values of $\tau$, over a discrete synthetic spectrum with nonuniform degeneracies $g(E_j)$. As $\tau$ increases, the filter sharpens and isolates individual energy levels. \textbf{Bottom:} The corresponding filtered partition function $Z_\tau(\lambda)$ becomes increasingly structured, developing valleys wherever $\lambda$ enters spectral gaps.}
    \label{fig:Z_sampling}
\end{figure}

If we compare this to the \textit{true} partition function
\begin{equation}
\label{eq:Zexact}
    Z_\tau(\lambda) = \text{Tr}\left[e^{-\tau(H - \lambda)^2}\right] = \sum_j g(E_j)\, e^{-\tau(E_j - \lambda)^2},
\end{equation}
an important distinction emerges. In the ITQDE correspondence $\tau$ sets the time evolution of each node via $\tau_k = \sqrt{\tau} x_k$ (with $x_k\in [-\sqrt{(m-1)},\sqrt{(m-1)}]$), it also acts as the width for a \textit{spectral filter}. $Z_\tau(\lambda)$ is after all nothing but a weighted sum of Gaussians, each centred on $E_j$. As shown in Fig.~\ref{fig:Z_sampling}, this allows $Z_\tau(\lambda)$ to be interpreted as the density of states convolved with a Gaussian filter of width $\frac{1}{\sqrt{\tau}}$. 

This filtering perspective presented in Fig.~\ref{fig:Z_sampling} shows $\tau$ directly controls the width of the filter being convolved with $g(E)$. This is consistent with the earlier identification of $\tau$ setting the resolution of the spectral staircase. 
This behaviour poses a hazard to the stability of any approximation to $Z_\tau(\lambda)$ however. Any gap  $\Delta_j \gtrsim \frac{1}{\sqrt{\tau}}$ will suppress $Z_\tau(\lambda)$ exponentially in the range $\lambda \in [E_j+ \frac{1}{\sqrt{\tau}},E_{j+1} -\frac{1}{\sqrt{\tau}}]$. This matters greatly, as it is relative to this minimum that the accuracy required by $\langle Z^{(\lambda)}_{\bar{m}} \rangle $ is determined. 

The potential instability induced by the decay of $Z_\tau(\lambda)$ is significant, as the \textit{relative} error from quadrature will be greatly amplified when the \textit{true} partition function $Z_\tau(\lambda)$ is on the order of this error. We can quantify this by appeal to Eq.\eqref{eq:quad-error}, where to leading order the truncation error corresponds to a threshold $Z_\varepsilon  \sim w_{\bar{m}+1} \sim \varepsilon$. Defining $\tau_\varepsilon=\tau_{\bar{m}+1}$, we may express the ITQDE approximated partition function  in terms of its true value and leading order residual contribution:
\begin{equation}
\label{eq:error_oscillation}
    \langle Z^{(\lambda)}_{\bar{m}} \rangle \approx Z_\tau(\lambda) + Z_\varepsilon \textrm{Re}[{\rm e}^{2i\lambda \tau_\varepsilon}].
\end{equation}

\begin{figure}[t]
    \centering
\includegraphics[width=1.05\linewidth]{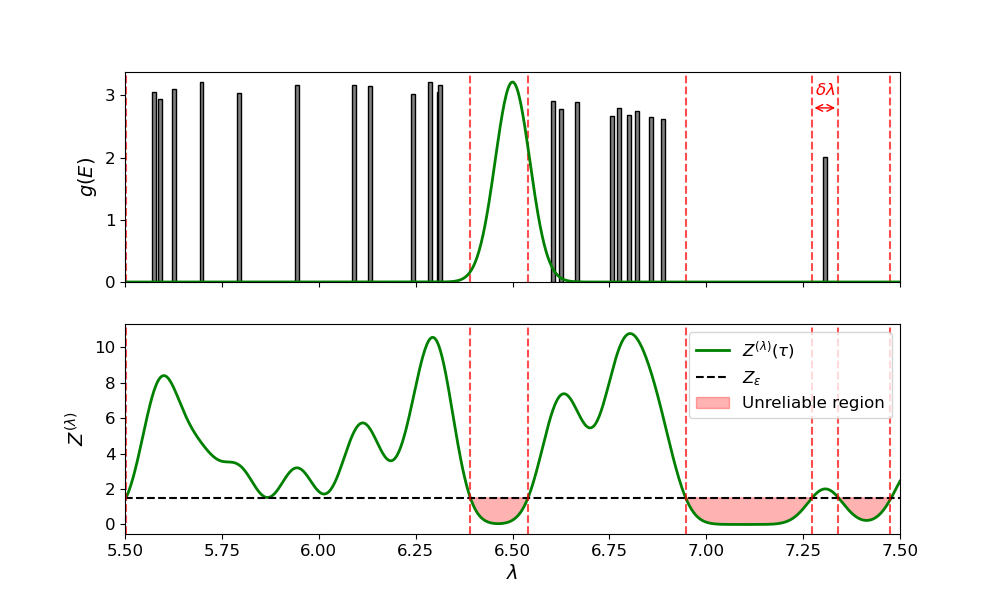}
    \caption{\textbf{Top:} Discrete synthetic spectrum with degeneracies $g(E_j)$. Red dashed lines denote boundaries of regions in which $Z_\tau(\lambda)$ with threshold $Z_\varepsilon$, with vertical spacing $\delta\lambda$ indicating the minimum width around which a single level can be resolved. \textbf{Bottom:} The filtered partition function $Z_\tau(\lambda)$ with threshold $Z_\varepsilon$ (dashed black). Regions where $Z{(\lambda)} < Z_\varepsilon$ are shaded red and denote unreliable output from the ITQDE reconstruction. These arise due to insufficient spectral support and cause the truncated quadrature tail to dominate the expectation estimate.}
    \label{fig:Z_floor}
\end{figure}

The analogous residual on the numerator $N_\varepsilon$ is bounded in relation to $Z_\varepsilon$ via $N_\varepsilon \le \|H\|Z_\varepsilon$. From this we may readily bound the approximation error of $\langle H_{\bar m}^{(\lambda)}\rangle$ to the true staircase $H_\tau (\lambda)$. To do so we define a \textit{local} tail ratio between the truncation threshold and the true partition function    $r(\lambda):=\frac{Z_\varepsilon}{Z_\tau(\lambda)}$, such that
\begin{equation}
\label{eq:minimal-bound}
\big|\langle H_{\bar m}^{(\lambda)}\rangle - H_\tau(\lambda)\big|
\le
 2\|H\| r(\lambda)+O\bigl(r(\lambda)^2\bigr).
\end{equation}

The accuracy of ITQDE is therefore dependent on the behaviour of $r(\lambda)$. To analyse this, we first let $d(\lambda):=\min_j |\,\lambda-E_j\,|$ be the distance from $\lambda$ to the nearest level. Then, further defining the local degeneracy via
\begin{equation}
    C_\ast(\lambda):=\sum_{j:|E_j-\lambda|=d(\lambda)} g(E_j), 
    \end{equation}we may bound $Z_\tau (\lambda)$ as
\begin{equation}
Z_\tau(\lambda)=\sum_j g(E_j)\,e^{-\tau(\lambda-E_j)^2}
\;\ge\; C_\ast(\lambda)\,e^{-\tau d(\lambda)^2},
\end{equation}
which in turn implies
\begin{equation}
\label{eq:r-lambda}
r(\lambda)\;\le\;\frac{Z_\varepsilon}{C_\ast(\lambda)}\,e^{+\tau d(\lambda)^2}.
\end{equation}
Clearly the behaviour of this tail ratio is exponentially dependent on the closest level to $\lambda$. In the worst case the local degeneracy is of order unity, and henceforth we take ${C_\ast(\lambda)}\equiv 1$. This can then be used to fix a stability threshold. We define $r_0$ as the tolerance, i.e. $r(\lambda)\leq r_0$ such that the pointwise approximation error will be bounded by $\|H\|r_0$). The saturation of this bound will correspond to an effective \emph{resolvable} radius around each spectral level in $\lambda$:
\begin{equation}
\label{eq:d-eps}
d_\varepsilon(\tau,\bar m)\;=\;\sqrt{\frac{1}{\tau}\,\ln\!\Bigl(\frac{1}{r_0Z_\varepsilon}\Bigr)}\,.
\end{equation}
At the midpoint of a gap of width $\Delta$, we have $d(\lambda)=\Delta/2$, meaning that $d_\varepsilon(\tau,\bar m)$ sets the \textit{largest resolvable gap} $\Delta_\varepsilon$ according to 
\begin{equation}
\label{eq:Delta-eps}
\Delta_\varepsilon=2d_\varepsilon(\tau,\bar m)
\;=\;2\sqrt{\frac{1}{\tau}\,\ln\!\Bigl(\frac{1}{r_0Z_\varepsilon}\Bigr)}\,.
\end{equation}
Reinserting the Gauss-Hermite expression for the residual threshold $Z_\varepsilon\sim e^{-d\bar m}$ with $d\simeq 2$, we obtain the approximate scaling between $\bar{m}$, $\tau$ and the \textit{maximum} gap that ITQDE can stably approximate: 
\begin{equation}
\label{eq:Delta-eps-linear}
\Delta_\varepsilon(\tau,\bar m)\;\simeq\;2\sqrt{\frac{2\,\bar m-\ln r_0}{\tau}}
\implies
\bar m\;\gtrsim\;\frac{\tau}{8}\,\Delta_\varepsilon^2
\end{equation}
This result constitutes a complementary constraint on ITQDE where $\tau$ sets both the smallest \textit{and} largest gaps that can be resolved. Fig.~\ref{fig:Z_floor} visualises this relationship, in which $Z_\varepsilon$ sets the threshold at which $Z_\tau(\lambda)$ is too small to be stably approximated with the chosen parameters.

\begin{figure}[t]
    \centering
    \includegraphics[width=\linewidth]{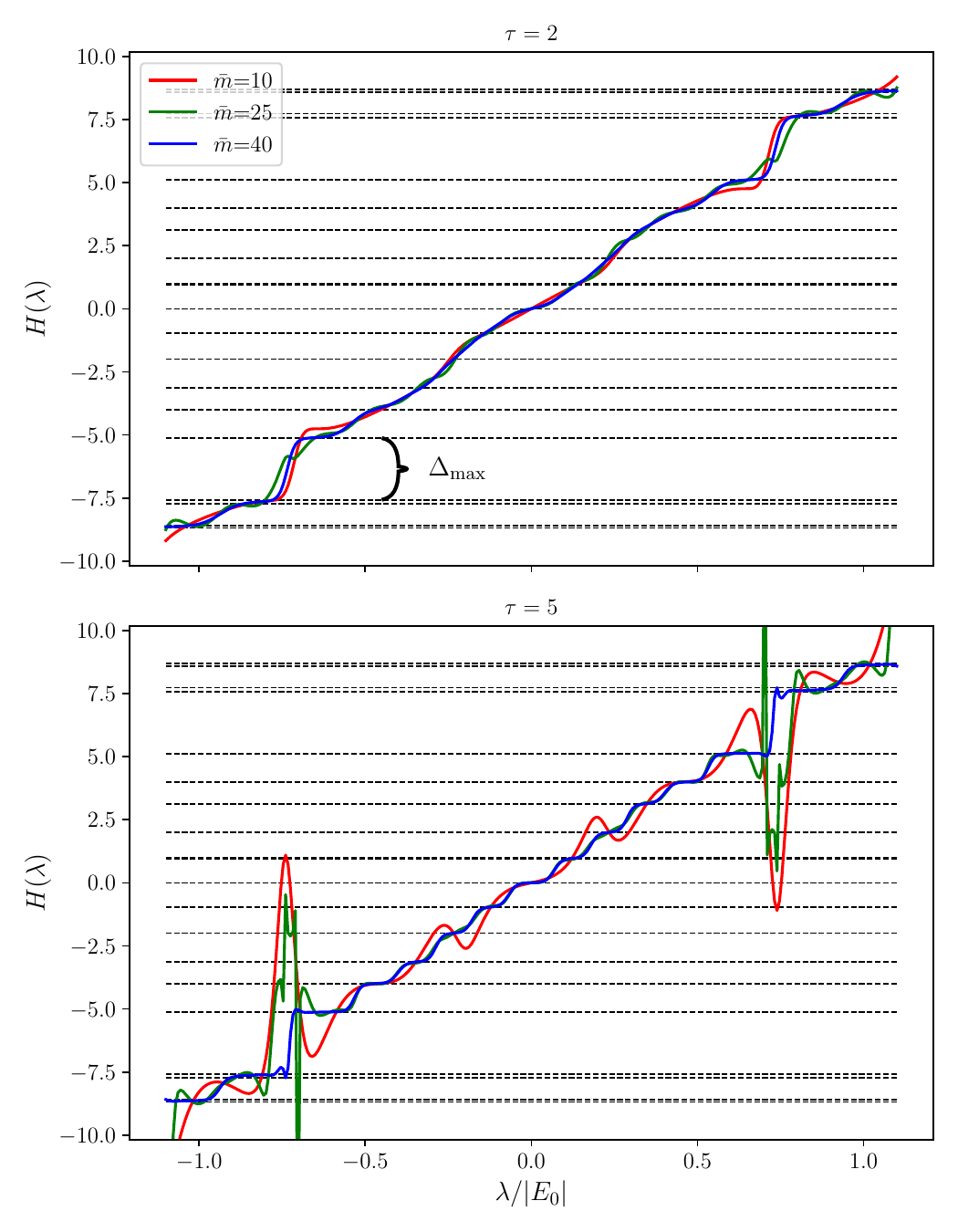}
    \caption{Illustration of oscillatory errors originating from the truncation to $\bar{m}$ overlaps, using the 2D-FH model using $U=2$,  $L_x=2$, $L_y =2$ and half-filling. Two values of $\tau$ are shown. In the top panel a more modest $\tau$ results in a smoothed staircase, where truncations to different $\bar{m}$ produce qualitatively similar curves.  Increasing $\tau$ to sharply resolve each level, we find a corresponding growth in the gap-dependent oscillations associated with truncation error. These are attenuated with larger $\bar{m}$.}
    \label{fig:ITQDEoscillations}
\end{figure}

Notably, the local nature of the gap-induced instability means it is readily identified. As Eq.\eqref{eq:error_oscillation} attests, the error from ITQDE is intrinsically oscillatory, and the presence of such oscillations in the staircase serve as a practical diagnostic of an over-resolved gap. This is readily demonstrated in Fig. ~\ref{fig:ITQDEoscillations} where the exponential loss of support from $Z_\tau (\lambda)$ induces spurious oscillations in the region of the largest gap. By varying both $\tau$ and the cut-off $\bar{m}$, we verify the behaviour predicted by Eq.\eqref{eq:Delta-eps-linear}. Namely that the oscillations around the largest gap $\Delta_{\rm max}$ are rapidly suppressed by increasing $\bar{m}$ or reducing $\tau$. 

Having completed this analysis, we return to the question motivating it — where lies the complexity in ITQDE's approximation of the spectral staircase? The remarkable convergence properties of quadrature suggest that both resolution and stability can be bought cheaply with linear scalings of parameters. Consequently the truncated ITQDE correspondence $\langle H_{\bar{m}}^{(\lambda)}\rangle$ appears to \textit{efficiently} approximate the true spectral staircase. Taking into account the scaling of the number of Hadamard tests required for estimating a $k$-local observable (see ref. \cite{leamer_quantum_2024}) we obtain that the overall scaling in the number of circuits required for a \textit{fixed} precision $\varepsilon$ will be $\sim {\rm poly}(n)$.

The tension here is that the QMA-hard nature of ground state energy estimation guarantees an exponential operational cost generically, and yet quadrature and truncation appear to have quashed this. In the next section, we shall see that the ultimate resolution to this lies in the \textit{sampling cost} of each overlap, rather than the number of overlaps required. 

\section{Sampling and Resolution Limits in ITQDE \label{sec:sampling}} 
We have seen that with quadrature, the ITQDE correspondence is exponentially accurate in $\bar m$ at only polynomial circuit cost (both in count and depth). Thus, for fixed $\tau$,
the deterministic floor $Z_\varepsilon$ can be made arbitrarily small with polynomial resources. What remains is the \emph{statistical} price of estimating the required expectations from finite shots.

Because the per-shot outcome on each overlap term is bounded, the variance on any estimators will follow the universal $\frac{1}{\sqrt{K}}$ behaviour, where $K$ is the total shot budget \cite{nielsen_quantum_2010}. In practice there are any number of ways to form the expectations required by ITQDE \cite{elben_cross-platform_2020,huang_predicting_2020, flammia_direct_2011, garttner_measuring_2017}, but the scaling of sampling cost with Hilbert-space dimension $D=2^n$ is best understood through the lens of stochastic trace estimation \cite{PhysRevA.80.012304}. This in effect treats a shot sample as the expectation on a random state, whose average will be proportional to the true trace. We reiterate that it is in the trace form of Eq.\eqref{eq:Z_trace} that we will consider ITQDE estimator quantities. 

The stochastic trace is performed by first preparing an initial state $\ket{\phi}$ drawn from either the Haar ensemble or any unitary/state $2$-design \cite{PhysRevA.80.012304}. The expectation of this state from $K$ samples  will be
\begin{equation}
\label{eq:haar-mean}
\mathbb{E}[\bra{\phi}O\ket{\phi}]= \frac{1}{K}\sum_k\bra{\phi_k}O\ket{\phi_k}=\; \frac{1}{D}\,\mathrm{Tr}[O],
\end{equation}
with variance
\begin{equation}
\label{eq:haar-var}
\mathrm{Var}\left(\bra{\phi}O\ket{\phi}\right)
= \frac{ \mathrm{Tr}(O^\dagger O) - \frac{|\mathrm{Tr}[O]|^2}{D}}{D(D+1)}.
\end{equation}
The \textit{relative} variance will then be
\begin{equation}
\label{eq:haar-var-rel}
\frac{\mathrm{Var}\left(\bra{\phi}O\ket{\phi}\right)}{\mathbb{E}[\bra{\phi}O\ket{\phi}]^2} =\frac{D}{D+1}
 \left(\frac{\mathrm{Tr}\left[O^\dagger O\right]}{\left[\mathrm{Tr}[O]\right]^2}-\frac{1}{D}\right).
\end{equation}
Under the mild assumption of $\mathrm{Tr}[O] \propto D$, this relative variance will shrink with $\mathcal{O}(\frac{1}{D})$,  shows that random state sampling will concentrate exponentially to its mean with system size $n$. There is however the competing behaviour of the \textit{orthogonality catastrophe} \cite{PhysRevLett.18.1049}, where the expected overlap between states vanishes with Hilbert space dimension. That is, while the relative variance of a random state expectation is exponentially suppressed, the $\frac{1}{D}$ factor means the \textit{same} argument applies to Eq.~\eqref{eq:haar-mean}. We are saved from this issue by the ratio form of the ITQDE calculation $\frac{\langle N^{(\lambda)}_{\bar{m}}\rangle}{ \langle Z^{(\lambda)}_{\bar{m}}\rangle}$. This cancels the leading order $\frac{1}{D}$ suppression, and if both numerator and denominator are estimated from the same ensemble of random states, their fluctuations are positively correlated. This covariance further reduces the variance of the ratio \cite{PhysRevLett.125.200501}  such that the relative precision is governed by the sampling shot budget $K$ and not by the Hilbert–space dimension.

A full analysis of the sampling error is performed in  App. \ref{app:Sampling}, with the principal result being that the spectral staircase's sampling error is bounded by:
\begin{equation}
\label{eq:minimal-bound-sampling}
\big|\langle H_{\bar m}^{(\lambda)}\rangle -H_\tau(\lambda) \big| \le \frac{H_\tau(\lambda)}{\sqrt{K}}\mathcal{O}(n^k).
\end{equation}
The polynomial $\mathcal{O}(n^k)$ factor emerges as a consequence of the bandwidth of a $k$-local Hamiltonian, and only meaningfully contributes in the bulk of the spectrum. At the spectral edge this can be replaced by a term of $\mathcal{O}(1)$. All other scalings with $n$ are benign and do not contribute to the sampling error bound.  In other words, while orthogonality–catastrophe effects shrink absolute estimates, the ITQDE ratio remains well–conditioned: the relative error obeys the universal statistical $\frac{1}{\sqrt{K}}$ law, independent of system size. 

\begin{figure}
    \centering
    \includegraphics[width=\linewidth]{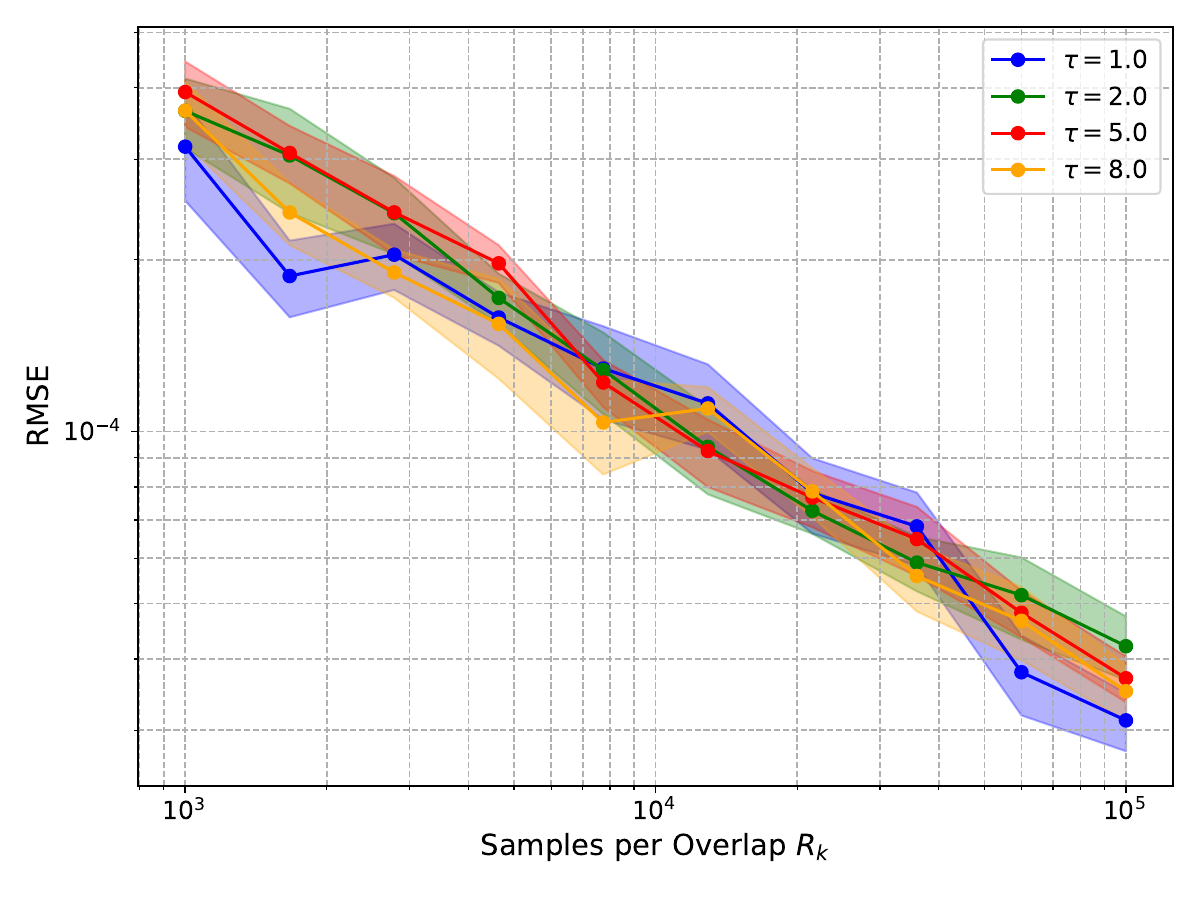}
    \caption{Error scaling of the ITQDE when estimated with stochastic trace estimation, applied to the 2D-FH model using $U=2$,  $L_x=2$, $L_y =3$ and half-filling. The overall error from sampling is calculated using Eq. \eqref{eq:error}, for four values of $\tau$ (with correspondingly chosen $m$ and $\bar{m}$). The shaded regions show the standard deviation of the error with respect to sampling. As we might expect, regardless of the value of $\tau$ (with sufficient $\bar{m}$), the error induced by sampling is fixed and follows the expected $K^{-\frac{1}{2}}$ scaling.}
    \label{fig:samplingscaling}
\end{figure}

A simple implementation of the stochastic trace is illustrated in Fig. \ref{fig:samplingscaling}, where expectations are taken with respect to a set of $K$ uniformly random $\{\ket{\phi_j}\}$ states. The trace can then be estimated from the mean expectation of these random states. The error between the true staircase and the stochastically estimated ITQDE is then calculated according to  Eq.\eqref{eq:error}. Using this proxy we observe the expected $\frac{1}{\sqrt{K}}$ scaling from sampling.

It is important to note that Eq.\eqref{eq:minimal-bound-sampling} is derived on the presumption that the relative error on the partition function due to sampling is small enough that the staircase ratio can be expanded in its errors. This corresponds to a threshold on the partition function $Z_{\varepsilon}^{\mathrm{(samp)}} \sim \frac{1}{\sqrt{K}}$. In other words the shot budget $K$ plays an identical role to the quadrature truncation threshold, insofar as it will set a resolvable radius, and — just as in Eq. \eqref{eq:Delta-eps} — a \textit{largest resolvable gap}.  Presuming $Z_{\varepsilon}^{\mathrm{(samp)}} \geq Z_{\varepsilon}$, this will depend directly on $K$ rather than $\bar{m}$. The passage from a deterministic quadrature truncation to statistical estimation error is therefore purely formal: the analysis of Sec.~\ref{sec:quadrature_stab} holds \textit{mutatis mutandis} under the substitution $Z_\varepsilon \to Z_{\varepsilon}^{\mathrm{(samp)}}$. In complete analogy with Eq. \eqref{eq:Delta-eps}, we then find the sampling dependence on the maximal resolvable gap scales as 
\begin{equation}
\label{eq:Delta-eps-K}
\Delta_\varepsilon\sim \sqrt{\frac{\ln(K)}{2\tau}}.
\end{equation}

It is this relationship through which the complexity barrier for energy estimation manifests itself. Let $\Delta_{\min}\!\sim\!\tau^{-1/2}$ denote the smallest gap resolvable at filter width $\tau$, and let $\Delta_\varepsilon$ be the stability threshold set by quadrature/sampling. ITQDE can stably reconstruct staircase steps only within
\begin{equation}
\Delta_{\min}\ \lesssim\ \Delta_j\ \lesssim\ \Delta_\varepsilon.
\end{equation}
Define the local gap bandwidth $\kappa_{\rm loc}=\Delta_{\max}/\Delta_{\min}$ over the spectral window of interest. When $\tau$ and $K$ are chosen so that both the smallest and largest gaps in that window are addressed (i.e., $\Delta_\varepsilon \gtrsim \Delta_{\max}$), the required shot budget inherits a scheme-specific exponential scaling through this local bandwidth:
\begin{equation}
\label{eq:K-bound}
K\ \gtrsim\ \exp\!\left[c\,\frac{\Delta_{\max}^{2}}{\Delta_{\min}^{2}}\right]
\;=\; \exp\!\big[c\,\kappa_{\rm loc}^{2}\big],
\end{equation}
with $c=O(1)$ determined by estimator constants and the stability margin. Operationally, the “QMA cost” manifests as the price of \emph{simultaneously} resolving the smallest gap while remaining stable in the presence of the largest gap in the same local window.

This has important implications for ground-state certification. Driving $\lambda$ below the spectrum until the (truncated) partition function crosses its threshold certifies only that any missing level lies at least $\Delta_\varepsilon$ beneath the current ground-state estimate; meanwhile the pointwise level resolution is set by $1/\sqrt{\tau}$. Together these two scales delimit (i) how far below an estimated level one may certify emptiness and (ii) how precisely that level can be estimated. Their ratio controls the exponential dependence of $K$ in Eq.~\eqref{eq:K-bound}. 

In the context of the ground-state, the two–scale obstruction—resolution set by $1/\sqrt{\tau}$ versus emptiness–certification depth $\Delta_\varepsilon$—pinpoints where costs blow up: when a local window contains both very small and comparatively large gaps, Eq.~\eqref{eq:K-bound} gives
$K \gtrsim \exp\!\big[c\,(\Delta_{\max}/\Delta_{\min})^{2}\big]$ for our ITQDE–quadrature scheme. This echoes the mechanism behind spectral–gap \emph{undecidability} in the thermodynamic limit: there exist translationally invariant lattice families whose gap behavior remains indistinguishable up to uncomputably large sizes before switching to a dense, gapless spectrum \cite{cubitt2015undecidability,bausch_undecidability_2020}. 
We do \emph{not} claim undecidability or a universal lower bound for spectral estimation here; rather, our $\kappa_{\rm loc}=\Delta_{\max}/\Delta_{\min}$ scaling shows how the same local coexistence of scales that defeats uniform certification in the undecidable constructions manifests concretely as a controllable resource divergence in a practical, finite–size estimator.

With this we may conclude efficiently resolving the smallest and largest gaps in a spectrum \emph{simultaneously} is precluded, and QMA is preserved. The instability causing this is however \emph{local} in the free parameter $\lambda$, and arises from the interplay of $K$ and $\tau$. Most importantly, this perspective reveals \textit{controllable trade-offs}. Those regions where exponential sampling is required are precisely those in which $\tau$ produces an unnecessarily fine filter relative to the gap size. These are readily identified due to the spurious oscillations they produce, and can be addressed with a downscaling of $\tau$ to the appropriate gap size. This then motivates the introduction of a controlled smoothing procedure in ITQDE. A classical procedure to \textit{effectively} rescale $\tau$ without re-computing the constituent estimators would then allow resolution to be traded for lower sampling costs. 

We may formalise this intuition as follows. Let $W_{\delta\lambda}$ be any even, unit-area window of width $\delta\lambda/\sqrt{2}$ and define the smoothed estimator
\begin{equation}
\widetilde{F}(\lambda)
=\int W_{\delta\lambda}(\lambda-\lambda')\,F(\lambda')\,d\lambda'.
\end{equation}
Convolving the Gaussian-filtered quantities (e.g.\ $N_\tau(\lambda)$) with a symmetric window is equivalent to replacing $\tau$ by an \emph{effective} value
\begin{equation}
\tau_{\mathrm{eff}}
=\frac{\tau}{\,1+\tau(\delta \lambda)^2\,},
\end{equation}
so that when $\delta\lambda \gg 1/\sqrt{\tau}$ the operative length scale becomes $1/\sqrt{\tau_{\mathrm{eff}}}\approx \delta\lambda$ and, to leading order (exactly for Gaussian $W$), $\widetilde{H}_\tau(\lambda)\simeq H_{\tau_{\mathrm{eff}}}(\lambda)$. In this sense, smoothing is a post-hoc, \emph{local} rescaling of $\tau$ that prevents over-resolution of gaps that cannot be stably sampled.

From the perspective of ITQDE's sum of oscillatory modes, smoothing multiplies a mode of local angular frequency $\omega(\lambda)$ by the factor $\widehat{W}_{\delta\lambda}(\omega)$ (for a Gaussian, $\exp[-\tfrac12(\delta\lambda\,\omega)^2]$), thereby suppressing high-frequency sampling noise in a way that is functionally equivalent to the $\tau_{\mathrm{eff}}$ map. 

\begin{figure}
    \centering
    \includegraphics[width=\linewidth]{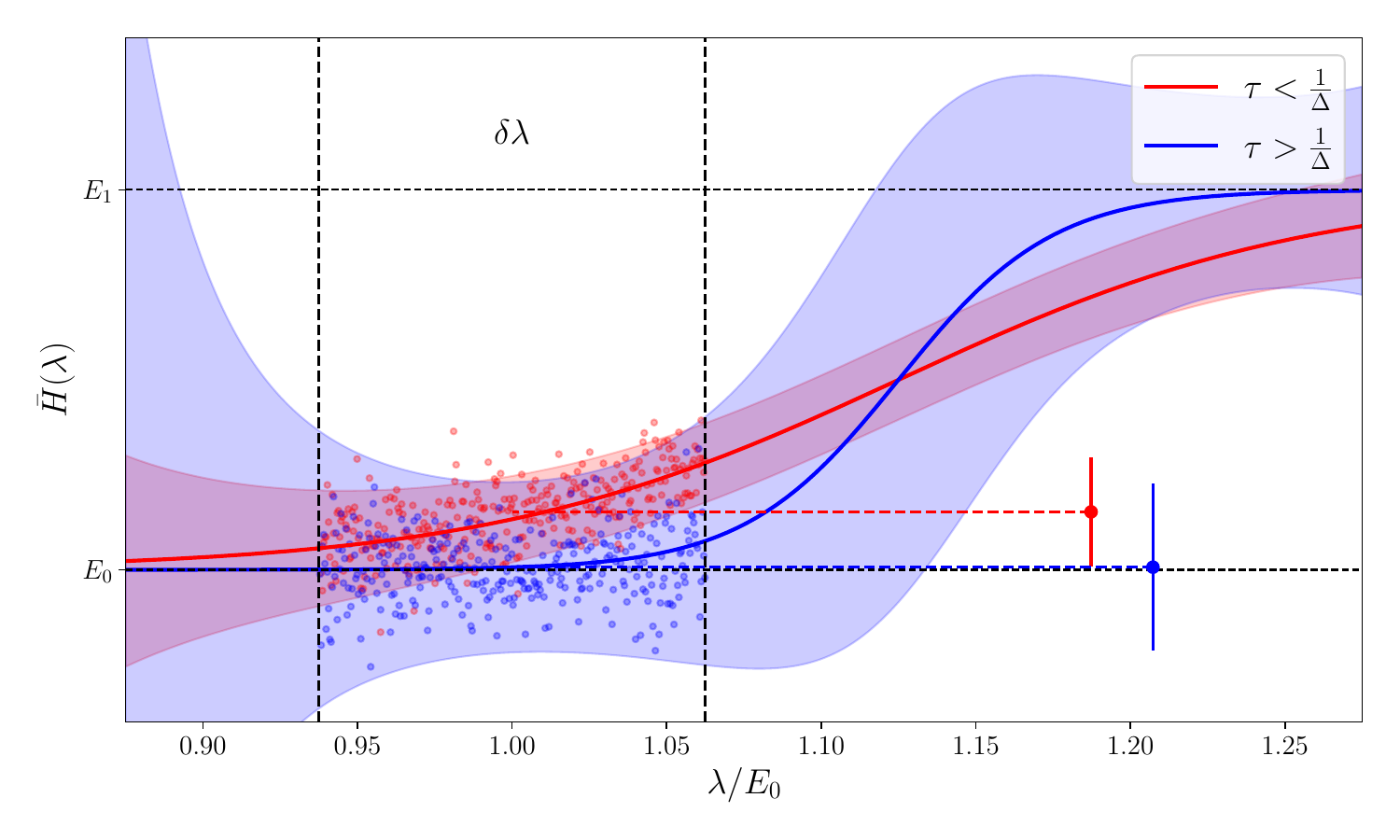}
    \caption{Resolution floor set by the trade-off between stability and spectral resolution.  Here we illustrate the sampled measurements using two values of $\tau$. The true value is plotted, together with the envelope of error due to sampling variance. In this example there are two levels $E_0$, and a neighbouring level $E_1 = E_0+\Delta$. Below $E_0$ there is no level to support the filter, and as expected this lack of support leads to an exponential growth in the variance envelope. To avoid instability from vanishing partition support, estimates must be smoothed over a window $\delta\lambda$.  The mean and variance from averaging over the window for both values of $\tau$ are also shown, highlighting the precision/bias trade-off set by the choice of $\tau$ relative to the spectral gap. A larger $\tau$ will give a more precise estimate of the energy level, at the price of increased variance, while a smaller $\tau$ yields more efficient sampling but a stronger bias in its estimate.}
    \label{fig:bias_precision}
\end{figure}

 Notably, because $\lambda$ is itself controlled only by a classical parameter, this scaling can be performed \textit{locally}. The ability to rescale $\tau$ in this post-hoc manner gives one access to a further bias–variance trade-off. This is shown in Fig. \ref{fig:bias_precision}, where we consider averaging over a window $\delta \lambda$  in a region containing two levels separated by a $\Delta$, for two underlying values of $\tau$. 

When $\tau>1/\Delta^2$ (blue), spurious oscillations from sampling error will grow exponentially away from the true value. Smoothing over $\delta \lambda$ will suppress the effective size of these oscillations (and therefore the sampling cost) by a factor exponential in $\frac{\delta \lambda}{\sqrt{\tau}}$. This smoothing stabilisation does not come without a cost however. When \(\tau<1/\Delta^2\) (red), the bare Gaussian kernel is already broader than the gap $\Delta$. This leads to both numerator and denominator having non-trivial spectral support over disjoint levels, and a systematic bias appears. A convenient bound for that bias is the Gaussian weight on the
far side of the gap:
\begin{equation}
\label{eq:bias-core}
|\phi_b(\lambda)|\;\lesssim\;\exp\!\big[-\,\tau_{\mathrm{eff}}\,\Delta^2\big],
\end{equation}
which can be derived directly from Eq. \eqref{eq:logistic}. As might be expected, this bias becomes relevant only for levels separated by a gap smaller than the coarse resolution set by $\delta \lambda$, and corresponds to the averaging out of distinct levels not resolvable on the scale of the smoothing filter.

Finally, the ability to rescale $\tau$ as a classical procedure means that the \textit{local} spectrum can always be obtained via coarse-graining. The original choice of $\tau$ then sets an upper resolution, but the ability to \textit{locally} rescale the spectrum classically means that \textit{the resolution can be reduced without cost}. For any over-resolved gap $\Delta_j$ we may choose to coarse-grain to:
\begin{equation}
\tau_{\rm eff}\;\sim\;\frac{\alpha}{\Delta_j^2}
\end{equation}
so that the effective kernel width matches the target gap scale. Notice that it is $\tau_{\rm eff}$ that sets $\Delta_{\rm min}=\frac{1}{\sqrt{\alpha}}\Delta_j$, while locally we may say  $\Delta_{\rm max}=\Delta_j$. Saturating the bound in Eq.\eqref{eq:K-bound} sets $\alpha \approx \ln(K)$. The \textit{gap dependent} resolution $\delta_j$ around the energy \(E_j\) under a shot budget $K$ will therefore be  
\begin{equation}
    \delta_j \approx \frac{1}{\sqrt{\ln(K)}}\Delta_j.
\end{equation}
Now, let us fix the shot budget to be $K\sim \mathrm{poly}(n)$. Provided the degree of this polynomial is greater than $2k$, then the error due to Eq.\eqref{eq:minimal-bound-sampling} will scale with $\frac{1}{\mathrm{poly}(n)}$. The bias under this coarse-graining will also be bounded by $\phi_b \le \frac{1}{K}$ and so inherit the same polynomial scaling. The bound on precision due to these sources thus matches the promise gap in the local Hamiltonian problem \cite{kitaev_classical_2002}. The critical difference lies in the fact that the overall precision on the ITQDE estimate will also depend on $\delta_j$. This will scale with $\frac{1}{\sqrt{\ln(n)}}$, but as a proportion of the local gap $\Delta_j$. For gapped systems where $\lim_{n\to \infty}\Delta_0=\Delta_*$, the achievable precision decouples from $n$. In this case to achieve a precision $\delta_j < \varepsilon$ would require $\ln(K)\lesssim (\frac{\Delta_*}{\varepsilon})^2$.    

The significance of this is worth emphasising. The QMA-hard problem (which may require exponential resources with $n$) is to estimate $E_0$ to precision $\frac{1}{\mathrm{poly}(n)}$. On the other hand, if ITQDE is restricted to $\mathrm{poly}(n)$ resources, we find that its estimation $E_j$ (and other eigenenergies) has a precision scaling with $\sim\frac{1}{\sqrt{\ln(n)}}$, but is \textit{multiplicative relative to the local gap}. In this sense, ITQDE occupies a pragmatic middle ground: it cannot generically evade the QMA-hardness of exact state preparation, but it does allow for efficient access to smoothed spectral information, with sampling costs that grow only polynomially in system size. This is the \textit{free snack} — an approximate solution that can be obtained at a relatively trivial cost to the full meal.

\section{Outlook \label{sec:Outlook}}
This work does not alter the hardness of ground-state estimation—the Local Hamiltonian problem remains QMA-complete. ITQDE does not promise a free lunch; it supplies a controlled, finite-resolution \textit{free snack}.  A profitable lens through which to view this result is through \textit{universality}: Eq.\eqref{eq:K-bound} ties sampling cost to a scale-free \textit{dimensionless} spectral bandwidth. Critically however, the absolute scale is chosen by hand with $\tau$. In this sense the smoothing operation is highly reminiscent of renormalisation group coarse-graining \cite{wilsonkogut1974}. Smoothing from $\tau \to \tau_{\rm eff}$ is analogous to a lowering of the UV cutoff, and acts as a one-way downsampling to a scale where the spectrum is locally tractable around $\lambda$. This permits the calculation of spectra and partition functions with resources that scale polynomially in system size and in the chosen filter width. While this entails a necessary trade-off between shot budget and resolution, in practical terms this is often sufficient: one does not need exact eigenstates to extract physical observables, thermodynamic properties, or even phase structure.  

What is striking is that this trade-off is enabled by the ability to identify where spectral estimation becomes \textit{locally} hard, and to rescale the problem to lower its effective cost. Shot budget sets an implicit resolution limit relative to $\tau$. Empirically, estimates follow the universal $K^{-1/2}$ sampling law, but the achievable bandwidth between resolution floor and ceiling improves only logarithmically with \$K\$. Oscillatory diagnostics can identify when a window is over-resolved, and coarsening to a smaller effective width $\tau_{\rm eff}$  stabilizes estimates by trading sampling variance for resolution in a controlled way (see Sec.\ref{sec:sampling}). This capacity may inform the future development of both adiabatic algorithms and imaginary-time schemes, where the role of spectral gaps and filter functions is central. It is also likely to be applicable to condensed matter and chemistry, where coarse spectral data are often enough to separate phases by locating plateaux and edges, estimate free energies via reweighting, and condition variational ans{\"a}tze with certified “empty-below” windows.

More broadly, since solutions to NP problems can be encoded into Ising-model ground states \cite{lucas_ising_2014}, approximate solutions to such tasks can in principle be addressed at finite resolution via ITQDE. The relatively minimal resources required by ITQDE, and existing proof-of-principle hardware implementations \cite{leamer_quantum_2024} are strongly suggestive that it can be applied on near-term NISQ hardware. The underlying QDE framework is transferable to other problems involving non-unitary operations; for example, it extends to more general forms of dynamics, including those generated by non-Hermitian Hamiltonians \cite{ivaki_dynamical_2025} and dissipative Lindbladian evolution \cite{McCaulensemble}. This points to a broader toolbox of NISQ-friendly probes operating below worst-case complexity barriers while still providing practical calculational utility.

Finally, the results here underscore a broader lesson: between the trivial and the intractable lies an operational middle ground. ITQDE makes spectral estimation the working language of that middle ground—delivering coarse answers today and reserving exponentially costly precision for when it is truly necessary.

 \begin{acknowledgments}
 This work was supported by the European Union and the European Innovation Council through the Horizon Europe projects QRC-4-ESP (Grant Agreement no. 101129663) and QUEST (Grant Agreement No. 101156088), in addition to the (UKRI) Horizon Europe guarantee scheme for the projects QRC-4-ESP  (Grant No. 10108296) and QUEST (Grant No. 10130220).  This work was made possible only by the assistance of Sarah and Sally Cavendish. Their contribution has been critical, although neither of them are physicists, and one of them is a dog. Finally, GM gratefully acknowledges Denys Bondar and Alicia Magann, not only for their reliably insightful suggestions, but their enduring friendship and support. He has needed every bit. 
\end{acknowledgments}

\appendix

\section{Quadrature approximated ITQDE}\label{app:quad_approx}
Here we derive the quadrature approximated ITQDE, and establish a bound on its error. This is achieved by considering limit of the correspondence obtained in Ref.\cite{leamer_quantum_2024}. Let us evaluate the integral via Gauss-Hermite quadrature (see Eqs. (3.5.15), (3.5.19), and (3.5.28) of Ref.~\cite{NIST:DLMF}),
\begin{align}
    & \frac{1}{2}e^{-\tau H^2} \notag\\
    & = \frac{1}{\sqrt{\pi}} \sum_m \int_{-\infty}^{\infty} e^{-x^2} \mathrm{Re}(e^{-2i\sqrt{\tau}x E_m}) \ket{E_m}\bra{E_m} dx \notag\\
    & = \frac{1}{\sqrt{\pi}} \sum_m \sum_{k=1}^K w_k \mathrm{Re}(e^{-2i\sqrt{\tau}x_k E_m}) \ket{E_m}\bra{E_m} + {\mathcal{E}} \notag\\
    & = \frac{1}{\sqrt{\pi}} \sum_{k=1}^K w_k \mathrm{Re}(e^{-2i\sqrt{\tau}x_k H}) + {\mathcal{E}},
\end{align}
where the size $K$ coordinate grid $\{x_k\}$ and weights $\{ w_k \}$ are tabulated in Ref.~\cite{NIST:DLMF}. Moreover, there exists reals $\{ \xi_m \}$ such that the error term is
\begin{align}
    {\mathcal{E}} = \frac{K! 2^K}{(2K)!} (-i\sqrt{\tau})^{2K} H^{2K} \sum_{m}   \mathrm{Re}(e^{-2i\sqrt{\tau} \xi_m E_m}) \ket{E_m}\bra{E_m}. \notag
\end{align}
To write it more compactly, we can say that there is a unitary operator ${\mathcal{U}}$ such that the error ${\mathcal{E}}$ is
\begin{align}
    {\mathcal{E}} = \frac{K! (2\tau)^K}{(2K)!} H^{2K}  {\mathcal{U}}.
\end{align}
where the error has been expressed in a polar operator form.

From Eq.~(5.6.1) of Ref.~\cite{NIST:DLMF}, we get
\begin{align}
    \sqrt{\frac{2\pi}{n+1}} \left(\frac{n + 1}{e}\right)^{n+1} < n!
    < \sqrt{\frac{2\pi}{n+1}} \left(\frac{n + 1}{e}\right)^{n+1} e^{\frac{1}{12n + 12}}.
\end{align}
Hence, for a unitary invariant matrix norm $\| \cdot \|$, we obtain the bound
\begin{align}
    \| {\mathcal{E}} \| < \frac{(K + 1)^{K + 1/2}}{(2K + 1)^{2K + 1/2}}e^{\frac{1}{12K + 12}} (2\tau e)^K \| H^{2K} \|,
\end{align}
which as $K \to \infty$ asymptotically approaches
\begin{align}
    \| {\mathcal{E}} \| = \mathcal{O}\left( \left(\frac{\tau e}{2K}\right)^K \| H^{2K} \| \right)
\end{align}

\section{Sampling error scaling \label{app:Sampling}} 
Since the ITQDE correspondence can be made exponentially accurate for only a linear increase in circuit depth, we can assess the effect of sampling error on the staircase by considering the estimators for $N$ and $Z$ directly. Specifically, let us label the estimator obtained by sampling from the Gaussian kernels as

\begin{equation}
\widehat H^{(\lambda)} \;=\; \frac{\widehat N^{(\lambda)}}{\widehat Z^{(\lambda)}}\,,
\qquad
H^{(\lambda)} \;=\; \frac{\mathrm{Tr}\,N^{(\lambda)}}{\mathrm{Tr}\,Z^{(\lambda)}}\,,
\end{equation}
when \(N^{(\lambda)}\) and \(Z^{(\lambda)}\) are estimated from \(K\) random states drawn from a unitary/state 2-design \cite{PhysRevA.80.012304}, using the \emph{same} batch for numerator and denominator. For any operator \(O\),
\begin{align}
\mathbb{E}\!\left[\langle\phi|O|\phi\rangle\right] &= \frac{1}{D}\,\mathrm{Tr}\,O, 
\label{eq:A1}\\
\mathrm{Var}\!\left(\langle\phi|O|\phi\rangle\right) 
&= \frac{\mathrm{Tr}(O^\dagger O)-\frac{|\mathrm{Tr}\,O|^2}{D}}{D(D+1)}.
\label{eq:A2}
\end{align}
Denoting the average over \(K\) i.i.d.~states by
\begin{equation}
\overline O=\frac1K\sum_{k=1}^K \langle\phi_k|O|\phi_k\rangle,
\end{equation}
we have
\begin{equation}
\mathbb{E}[\overline O]=\frac{1}{D}\mathrm{Tr}\,O,\qquad
\mathrm{Var}[\overline O]=\frac{1}{K}\,\mathrm{Var}\left[\langle\phi|O|\phi\rangle\right].
\label{eq:A3}
\end{equation}
Now define $\overline N:=\overline{N^{(\lambda)}}$ and $\overline Z:=\overline{Z^{(\lambda)}}$
Using the 2-design second moment, the \emph{cross covariance} for a shared batch is
\begin{equation}
\mathrm{Cov}(\overline N,\overline Z)
=\frac{1}{K}\,\frac{ \mathrm{Tr}(N^\dagger Z)-\frac{\mathrm{Tr}N^\dagger\,\mathrm{Tr}Z}{D} }{D(D+1)}.
\label{eq:A4}
\end{equation}

\subsection*{Ratio variance (universal \(1/\sqrt{K}\) law)}

Write relative single–channel errors
\[
\varepsilon_N:=\frac{\overline N-\mu_N}{\mu_N},\quad
\varepsilon_Z:=\frac{\overline Z-\mu_Z}{\mu_Z},\quad
\mu_N=\frac{\mathrm{Tr}N}{D},\ \mu_Z=\frac{\mathrm{Tr}Z}{D}.
\]
A first–order expansion of \(\widehat H/H=(1+\varepsilon_N)/(1+\varepsilon_Z)\) yields
\[
\frac{\widehat H-H}{H}=\varepsilon_N-\varepsilon_Z+O(K^{-1}).
\]
From this we may immediately define the relative standard error on the staircase as
\begin{align}
\frac{\mathrm{Std}(\widehat H^{(\lambda)})}{H^{(\lambda)}}&=\sqrt{\mathrm{Var}(\varepsilon_N)+\mathrm{Var}(\varepsilon_Z)-2\,\mathrm{Cov}(\varepsilon_N,\varepsilon_Z)}
\notag\\
&=\sqrt{\frac{D}{D+1}}\ \frac{c(\lambda)}{\sqrt{K}} \;+\; O(K^{-1})
\label{eq:A5}
\end{align}
using
\begin{equation}
c(\lambda):=\sqrt{\frac{\mathrm{Tr}(N^\dagger N)}{|\mathrm{Tr}N|^2}
+\frac{\mathrm{Tr}(Z^\dagger Z)}{|\mathrm{Tr}Z|^2}
-2\,\frac{\mathrm{Tr}(N^\dagger Z)}{\mathrm{Tr}N^\dagger\,\mathrm{Tr}Z}}.
\label{eq:A6}
\end{equation}
If numerator and denominator use \emph{independent} random-state batches, the covariance term drops and \(c(\lambda)\) increases, but the \(1/\sqrt{K}\) scaling remains. If we now exploit the explicit form $N^{(\lambda)}=H Z^{(\lambda)}$, we can bound this term by Cauchy–Schwarz:
\begin{equation}
c(\lambda)^2
\;\le\;
\Bigg(\frac{\sqrt{\mathrm{Tr}(Z H^2 Z)}}{|\mathrm{Tr}(HZ)|}
      +\frac{\sqrt{\mathrm{Tr}(Z^2)}}{|\mathrm{Tr}Z|}\Bigg)^2.
\end{equation}
Moreover, $\mathrm{Tr}(Z H^2 Z)\le \|H\|^2\,\mathrm{Tr}(Z^2)$ and
$|\mathrm{Tr}(HZ)|=|H(\lambda)|\,|\mathrm{Tr}Z|$ with
$H(\lambda)=\mathrm{Tr}(HZ)/\mathrm{Tr}Z$. Hence
\begin{equation}
c(\lambda)\;\le\;\sqrt{Q(\lambda)}\left(\frac{\|H\|}{|H(\lambda)|}+1\right),
\end{equation}
where 
\begin{equation}
Q(\lambda):=\frac{\mathrm{Tr}(Z^2)}{|\mathrm{Tr}Z|^2}.
\end{equation}
Notice that depending on the width of the filter relative to the spectral bandwidth,  $\frac{1}{D} \leq Q(\lambda) \leq 1$. In this sense, the error due to sampling only has benign system size dependence. Note also that for a $k$-local Hamiltonian with $\|H\|=O(n^k)$ (hence spectral bandwidth
$\Delta S=O(n^k)$), the bound
\[
c(\lambda)\ \le\ \sqrt{Q(\lambda)}\Big(\tfrac{\|H\|}{|H(\lambda)|}+1\Big)
\]
yields:
(i) near band edges, $|H(\lambda)|=\mathcal{O}(\|H\|)$ so $c(\lambda)=\mathcal{O}(\sqrt{Q})$;
(ii) near the spectral center, $|H(\lambda)|=O(1)$ so
\begin{equation}
c(\lambda)=\mathcal{O}\big(n^k\sqrt{Q(\lambda)}\big).
\end{equation}
Thus the relative standard deviation obeys
\begin{equation}
\frac{\mathrm{Std}(\widehat H^{(\lambda)})}{H^{(\lambda)}}
\ \le\ \frac{1}{\sqrt{K}}\times
\begin{cases}
\mathcal{O}\big(\sqrt{Q(\lambda)}\big), & \text{edge},
\\ \mathcal{O}\big(n^k\sqrt{Q(\lambda)}\big), & \text{center}.
\end{cases}
\end{equation}
In bulk regions with a smooth density of states (on the scale) $\frac{1}{\sqrt{\tau}}$, $Q(\lambda)\approx
\sqrt{\tau}/\!\big(2\sqrt{\pi}\,D\,\rho(\lambda)\big)$, meaning relative error will drop exponentially with $n$. When levels are isolated $Q(\lambda)\approx 1$, and the \emph{polynomial} $n^k$ factor reappears in the worst case (central $\lambda$). From this we may state the final worst-case error from sampling with $K$ shots:
\begin{equation}
\label{eq:minimal-bound-sampling-app}
\big|\langle H_{\bar m}^{(\lambda)}\rangle -H_\tau(\lambda) \big| \le \frac{1}{\sqrt{K}}\mathcal{O}(n^k)
\end{equation}

Critically however, this argument relies on being able to expand the ratio $N/Z$ to linear order in their relative errors.  in In particular, the bound shows that the \emph{prefactor} $c(\lambda)$ is benign (at most
$O(\sqrt{Q})$) whenever $|\mathrm{Tr}Z|$ is not vanishing. The real constraint comes from the
\emph{linearisation} itself: it requires $|\varepsilon_Z|=\big|\tfrac{\overline Z-\mu_Z}{\mu_Z}\big|\ll1$.
With per–overlap sampling, $\overline Z$ is a weighted sum of bounded outcomes, so it has an
\emph{additive} sampling floor of order $K^{-1/2}$, essentially independent of the (possibly very
small) mean. In a gap, however, $|\mathrm{Tr}Z|$ decays exponentially with $\tau$ and the gap width;
at the midpoint of a gap $\Delta$ one has $|\mathrm{Tr}Z|\propto e^{-\tau\Delta^2/4}$. Thus
perturbativity requires
\[
e^{-\tau\Delta^2/4}\ \gg\ K^{-1/2}
\quad\Longleftrightarrow\quad
\Delta\ \lesssim\ \sqrt{\frac{\ln K}{2\,\tau}}.
\]
This bound on resolvable gap size is precisely analogous to that found via the truncation threshold in Sec.\ref{sec:quadrature_stab}, but here obtained through the analysis of sampling error on the spectral staircase.

\bibliography{literature.bib}

\end{document}